\documentstyle{elsart}
\begin{document}

\input epsf
%
%
\begin{frontmatter}
\title{
The Future of Mr. Jefferson's Laboratory (n\'ee CEBAF)\thanksref{talk}}
\thanks[talk]{Written version of an invited talk presented at
the Second ELFE Workshop on Hadronic Physics (St. Malo, France, September
1996).}
\author
 {Carl E. Carlson}
\address
 {Physics Department, The College of William and Mary, 
      Williamsburg, VA 23187, USA}
\begin{abstract}
We present one viewpoint plus some general information on
the plans for energy upgrades and physics research at the
Jefferson Laboratory.
%
%

PACS number(s): 13.60-r; 24.85+p 13.25Es 14.80.Ly
\end{abstract}
\end{frontmatter}
%
%
\section{Introduction}

This talk reports one viewpoint plus some general information on
the plans for energy upgrades and physics research at the
Jefferson Laboratory.  

The first thing to report is that the name of the place
has changed:  on 24 May 1996 the former Continuous Electron Beam
Accelerator Facility (CEBAF) became the Thomas Jefferson National
Accelerator Facility (TJNAF), or the Jefferson Laboratory or JLAB
colloquially.  Thomas Jefferson lived from 1743 to 1826, wrote the
Virginia Statute of Religious Freedom and the American Declaration of
Independence,  was the second Governor of Virginia under independence,
and the third President of the United States under the present
constitution. In Virginia he is still referred to as ``Mr.'' Jefferson.

The actual subject of the talk is the prospect for higher energy at
JLAB and what might be done with higher energy there.  Skipping over
politics, I will make a few comments about what I will call
engineering, meaning plans and hopes and methods for energy upgrades
and when they may occur, and then devote the bulk of the talk to physics.

\section{Engineering}

There are plans to upgrade the accelerator to what I am told should be
called $9 \pm 1$ GeV.  This can be done by (relatively) inexpensive
tuning of existing hardware and by adding to the accelerating
sections.  Getting energy beyond about 10 GeV would involve redoing the
bending arcs and would not be cheap.

The first step in the upgrade takes advantage of the fact that on the
average, the accelerating cavities have an accelerating gradient well
above specification.  Further, there is a factor of about 2 spread in
the accelerating gradients, and the reasons for this spread are becoming
better understood.  Thus there is the possibility and the plan that,
using the existing maintenance budget, the low gradient cavities can be
removed, refurbished, and then reinstalled as high accelerating gradient
cavities.

The tentative time frame is to have 6 GeV in 1997 and ``7+'' GeV in
1999.  (By the way, as of the time of this talk, TJNAF has already run
at 1 GeV/pass for one pass (the machine usually uses 5 passes or 5
circulations).)  If we talk of extra cost rather than absolute cost, the
extra cost of step one is zero, since the money is already in the
maintenance budget.  The likelihood of step one is quoted as at the ``90\%
confidence level.''

Step two involves increasing the number of accelerating cavities by
25\%, and put them in existing empty spaces in the straight section. 
This, incidentally, is possible because of a 1987 decision to have 5
passes instead of 4, reducing the number of accelerating cavities
needed.  But the concrete had already been poured, so that these empty
spaces exist.

The time frame for this step is to submit an accelerator upgrade
proposal in 1999, and then maybe in 2001 to have an energy of $9 \pm 1$
GeV.  The ballpark cost of this upgrade is \$20 million, and the
likelihood can only be guessed.


\section{Physics}


A useful, albeit two years old, source of information about
experiments at a higher energy CEBAF is the proceedings of the
``Workshop on CEBAF at Higher Energies'' that was held in April,
1994~\cite{workshop}.  The purpose of the workshop was to assess what
could be done at CEBAF with an 8 or 10 GeV electron beam.

The proceedings is organized into four headings, which I will also use
as my next four headings.  I will give some idea of the topics under
each heading, and study (from a 1996 perspective when updating is
relevant) in more detail a few examples of potential experiments.
I will also add one heading on a possible exotic use of TJNAF.


\subsection{Hadron spectroscopy and production}


There are a number of things to do with higher energy; some are ``more of
the same'' but better and others are new initiatives not possible at
lower energy.  Let us start with some numbers to illustrate with gains
follows from an energy upgrade.  We can work out that a 4 GeV electron
beam allows producing 

\hskip 1 in $\bullet$ baryons up to 2.9 GeV mass,

\hskip 1 in $\bullet$ mesons up to 2.0 GeV mass,

\hskip 1 in $\bullet$ strange mesons up to 1.8 GeV mass.

This seems fine for baryon spectroscopy studies, but a number of
interesting strange quark $s \bar s$ mesons states are predicted around 2
GeV, and these cannot be reached without more initial energy.  For
producing mesons
$M$ of mass $\mu$ in $\gamma^* + p \rightarrow M + B$, we get a limit
\begin{equation}
\mu \leq \sqrt{m_N \left( m_N + 2E_\gamma \right) - Q^2} - m_B
\end{equation}

\noindent (with $Q^2$ positive for spacelike photons).  Maximizing
$E_\gamma$ and minimizing $Q^2$, we with 8 (or 10) GeV incoming
electrons can produce

\hskip 1 in $\bullet$ baryons up to 4.0 (4.4) GeV mass,

\hskip 1 in $\bullet$ mesons up to 3.0 (3.5) GeV mass,

\hskip 1 in $\bullet$ strange mesons up to 2.9 (3.3) GeV mass,

\noindent enough to produce the $s \bar s$ we mentioned.

So it seems that even 8 GeV is enough to do the hidden-strange
meson spectroscopy we want to do.  In addition, one will want to
undertake hunts for glueballs ($gg$ or $ggg$ states), hybrids 
($q \bar q g$, in the meson version), oddballs ($J^{PC}=1^{-+}$
states, which are guaranteed not $q \bar q$), study
possibilities for a $\phi$ factory for $CP$ violation and rare
$\phi$ decay studies~\cite{dzierba}, and glueballinos.

\begin{figure}

\hskip 1. in  \epsfysize= 2.7 in   \epsfbox{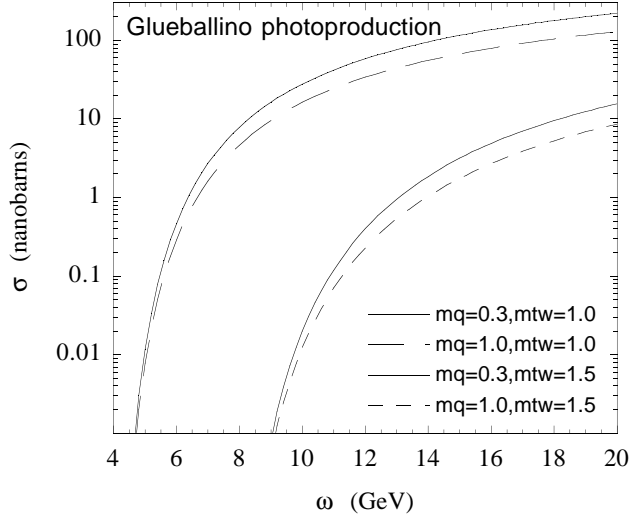}

\caption{Glueballino photoproduction cross section.  The upper
and lower curves are for glueballino masses (``mtw'') 1.0 GeV and
1.5 GeV, respectively.  The solid and dashed curves are for up or
down quark masses (``mq'') 300 MeV and 1 GeV, respectively. }
\label{gluino}

\end{figure}

The last are bound states of gluons and gluinos, with the latter
being the supersymmetric partner of the gluon and is clearly an
interesting particle to find.  The usual reason given for not
having found the gluino is that it is very heavy, but in fact
there is a window open for the possibility that the gluino is
light, under 2 GeV, if it is long lived, with a mean life of over
100 picoseconds~\cite{light}.  TJNAF has a good energy for producing
light gluinos---they will not be moving too fast in the final
state---and a good intensity.  Calculations of gluino production
are relatively easy since the vertices are just supersymmetric
partner of well known vertices, and the Feynman diagrams involve
only propagators of discovered (i.e., known quarks and gluons)
particles.  Fig.~\ref{gluino} shows a plot of cross section vs.
incoming photon energy for gluino photoproduction~\cite{gluino},  for
different guesses for the glueballino mass, and showing not much
sensitivity to what one puts in for the up or down quark mass.


Searching for glueballinos requires observing multiparticle final
states, hence is a Hall B experiment.   The luminosity in Hall B
is anticipated as 10 nb$^{-1}$sec$^{-1}$, so that there will be
quite a bit of glueballino production if the mass is 1 GeV.  the
value of extra energy is, however, clear.  The signatures for
glueballinos is like that for a weakly decaying meson: copious
production, slow decay (a significant gap between decay and
production point), and two much mass to be confused with a kaon. 
One possibility is to find $\pi^+\pi^-$ pairs with a mass above
the kaons, corresponding to the decay glueballino into 
$\pi^+\pi^- \tilde \gamma$ with $\tilde \gamma$ representing the
unobserved photino.  Finding the gluino is not necessarily to be
expected, but is not impossible, and would be a great
confirmation of an important hypothesis in particle physics and a
great {\it coup} for the machine that finds it.


\subsection{Exclusive reactions at high $Q^2$}


Possibilities under this heading begin with extensions of
things that have been begun at lower energies, particularly
including measurements of form factors of various hadronic
systems.  Meson form factors could be measured to $Q^2$ of 
GeV$^2$ if the beam is 8--10 GeV incoming electron energy. 
Separated measurements of baryon form factors, both elastic and
transition, could be done for $Q^2$ up to 10 GeV$^2$ if the
beam is 8 GeV.  (Incidentally, they can be done up to 7 GeV$^2$
even with a 4 GeV beam, according to the Proceedings of the
Higher Energy workshop.)  Few nucleon systems could also be
studied.  The $A$ and $B$ form factors of the deuteron could be
measured to $Q^2$ of 9 GeV$^2$ if the beam energy is 8 GeV,
compared to 6 GeV$^2$ for a beam energy of 4 GeV.  And the
deuteron photodisintegration reaction, 
$\gamma + d \rightarrow p + n$, could be measured to whatever
energy what available.  Remarkable measurements have already
been made up to $E_\gamma = 4$ GeV.

Further exclusive and semi-exclusive reactions can also be
studied.  The goal always is to learn something about the
structure of hadrons.  Take as an example the photoproduction of
high transverse momentum pions,  both from the viewpoint of
learning something about the pions and something about the
targets they are produced off.

Taking here the question of what we can learn about pions,
recall that exclusive reactions at sufficiently high momentum
transfers generally depend upon the distribution amplitude
$\phi$ of the particles involved, which in turn is the valence
quark wave function $\psi$ integrated over the momenta transverse
to the direction of the parent hadron.  For a pion,
\begin{equation}
\phi_\pi(x) = \int d^2k_T\, \psi_\pi(x,k_T)  .
\end{equation}

\noindent  A logarithmic scale dependence is tacit, and some
factors of 2 and $\pi$ have been ignored; $x$ is the
(light-front) momentum fraction carried by one of the quarks.  It
happens that both the pion electromagnetic form factor and the 
$\pi^0 \gamma \gamma$ form factor depend on the same integral,
\begin{equation}
I_\pi = \int dx\, {\phi_\pi(x) \over x} .
\end{equation}

\begin{figure}

\hskip 1. in  \epsfysize= 1.2 in   \epsfbox{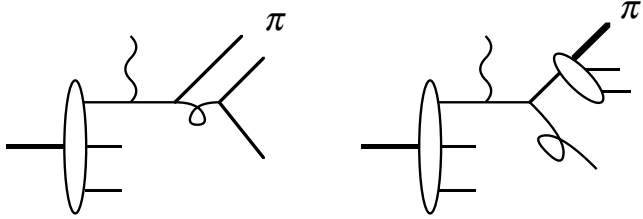}

\caption{Two ways to photoproduce pions.  The direct process on
the left dominates the the fragmentation process at high
transverse momentum, if the overall energy is not too large.}

\label{pion}

\end{figure}

One would like another way to measure this integral, and a way
may be provided by pion photoproduction.  The direct process,
Fig.~\ref{pion} left, has the pion produced in a short range
process before the relevant particles leave the immediate
reaction region. It is interesting here because the amplitude is
proportional to precisely the same integral, $I_\pi$.

\begin{figure}

\hskip .8 in  \epsfysize= 2.5 in   \epsfbox{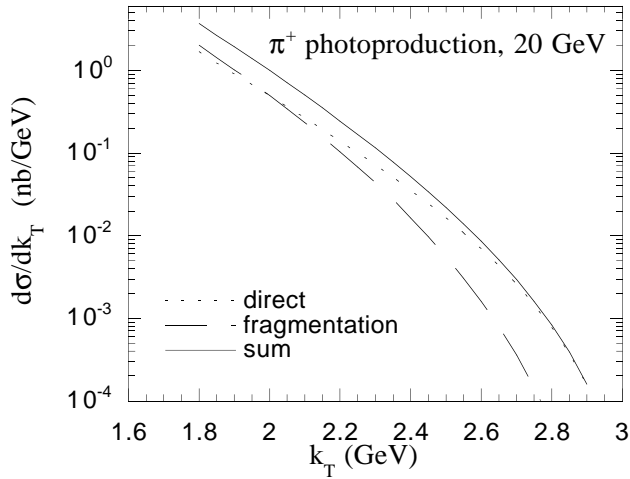}

\caption{Cross section for $\gamma + p \rightarrow \pi^+ + X$,
integrated over longitudinal momentum. (The calculation for
the direct process used the asymptotic pion distribution
amplitude.)}
\label{plot}
\vglue -10pt
\end{figure}

There is however a serious theoretical background, which is
production of high transverse momentum pions by producing
quarks or gluons moving fast in a particular direction, and
having those partons fragment into hadrons, one of which
becomes the observed pion, as in Fig.~\ref{pion} right.  At low
transverse momentum, fragmentation is indeed the dominant
process.  However, as the momentum gets higher, the likelihood
that a single pion can carry a large fraction of the partons
momentum decreases sufficiently that the direct process can be
seen.  A typical plot is shown in Fig.~\ref{plot}, for $\pi^+$
production with $E_\gamma$ = 20 GeV and integrated over longitudinal
momentum~\cite{wakely}.  


If data can be gotten at the higher transverse momenta, it will
be a direct measure of $I_\pi$ (modulo higher twist
corrections).  The calculations that went into Fig.~\ref{plot}
did assume that we knew the proton quark distributions, but
these we think are in fact decently well known now.  On the
other hand, observations of how the pion production rate
depends upon polarization of the beam and target depends on
polarized quark distributions, which are not well determined. 
Such observations may well be a flavor sensitive way to
learn something about the target's polarized quark
distributions.


\subsection{Inclusive and semi-exclusive scattering}


This section will be brief.  The proceedings of the Higher
Energy Workshop reports several good talks and
interesting experiments.  An example is the measurement of the
spin dependent structure function $g_1$ at higher $x$.  However,
one particular general problem for semi-exclusive processes was
that at the energies under discussion, it would be hard to make
a clear distinction between the target and current fragmentation
regions.  The summarizers of this topic reached the overall
conclusion that 8--10 GeV was too little energy~\cite{froismulders}.


\subsection{Hadrons in the nuclear medium}


Good experiments to learn things from the behavior of hadrons
in the nuclear medium include color transparency, vector meson
electroproduction off nuclei, and virtual Compton scattering off
nuclei.  One of the serious questions about the nucleus is what
are the right degrees of freedom to describe baryonic matter
when the baryons get very close together.  It may be that
nuclei can always be described as a collection of neutrons and
protons, and that repulsions in the nuclear force keep the
nucleons far enough apart so that the nucleons themselves keep
more or less the same character that they have in isolation. 
On the other hand, there may be changes in the matter, and one
possibility is that quarks in groups of six, in the simplest
cluster, reorganize into a new state.  If each quark is in the
lowest spatial state, the color-spin-flavor part of the wave
function is fixed by antisymmetry and the overall spin-isospin
quantum numbers.   We shall refer to this as the 6q model, and
one problem is how to tell 2N and 6q nuclear configurations apart
experimentally.  For many experiments, things like the wave
functions in the 2N model may be adjusted to give the same
results as one would expect from the 6q model.  For example,
Hanlon, Lassila, and I~\cite{hanlon} looked at the spectrum of backward
moving protons, $p_B$, in 
$\ell + d \rightarrow \ell' + p_B + X$, where $\ell$ and
$\ell'$ are leptons, and found that pure 2N with any of a
collection of standard wave functions gave calculated results
that were below the data above 400 MeV backward hemisphere
momentum.  Adding one or a few percent, by normalization, 6q
state allowed matching the data well.  However, this is not a
proof that the 6q state must be present because it is easy to
see how a modest (by the standards of the trade) increase in the
tail of the wave function in the 2N calculation would allow a
fit to the data.

We need a different type of suggestion to differentiate 2N and
6q contributions to the cross sections, for example, one that
takes advantage of how the cross section factorizes differently
in the two calculations.  For backward proton production in a
2N calculation, the process proceeds with the neutron being
struck, with its pieces going forward, and the proton emerging
with whatever fermi momentum it had when the neutron was
struck---backward in the cases of interest.  The cross section
is
\begin{equation}
{d\sigma_{2N} \over dx\, dy\, d\alpha\, dp_T} \equiv \sigma_{2N}
  =  K F_{2n}(\xi)(2-\alpha) |\psi(\alpha,p_T)|^2.
\end{equation}

\noindent Here, $x$ and $y$ are lepton variables, $Q^2/2m_N\nu$
and dimensionless lepton energy loss (in the lab, lepton energy
loss divided by incoming lepton energy), respectively.  Variables
$\alpha$ and $p_T$ give the momentum of the backward proton,
with $p_T$ being the momentum transverse to the incoming lepton
momentum and 
\begin{equation}
\alpha \equiv {E_p + p^z \over m_N}.
\end{equation}

\noindent $K$ is a known kinematic factor,  $\psi$ is the 2N
wave function of the deuteron, and the argument of the neutron
structure function $F_{2n}$ is the momentum fraction of the
struck quark relative to the momentum of the neutron.  If the
neutron were stationary, it would precisely $x$, but since the
neutron is moving,
\begin{equation}
\xi = {x \over 2-\alpha}.
\end{equation}

Now we can make a suggestion to distinguish the 2N and 6q
predictions by studying the two-nucleon test ratio~\cite{lassila},
\begin{equation}
R \equiv {\sigma_{meas} \over K F_{2n}}.
\end{equation}

\begin{figure}

\hskip 1.1 in  \epsfysize= 2.5 in   \epsfbox{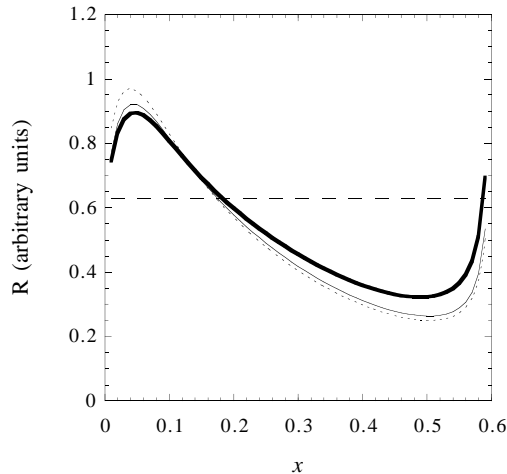}

\caption{The two nucleon test ratio $R$ for $p_T = 0$
and $\alpha = 1.4$, or 322 MeV backward protons.  The flat
dashed line is for 2N, the curves are for three related
implementations of a 6q cluster model.}

\label{R}

\end{figure}

If the measured cross section is due to a 2N configuration, 
$\sigma_{meas}=\sigma_{2N}$, then the ratio $R$ is independent
of the lepton variables $x$ and $y$.  One could, for fixed $p_T$
and $\alpha$ and integrating over $y$ (both numerator and
denominator, before taking the ratio), simply plot $R$ versus
$x$.  The result should be a flat line, as illustrated by the
dashed line in Fig.~\ref{R}.


In advance of enough data to make such a plot the question for
a theorist is, how different is the 6q prediction?  The answer,
based on some reasonable modeling for the quark distributions
in the 6q cluster and some use of counting rules to get the
backward proton spectrum, is also shown in Fig.~\ref{R}. Of
course, the deuteron is  not 100\% 6q cluster, but if we look
only at high backward momentum protons, we enhance the fraction
of events that come from when the nuclear matter is all close
together at the outset, and hence could expect a large fraction
of the observed protons to come from a 6q cluster. Hence, if a 6q
cluster is present even with small overall normalization, many
of the fast backward protons could come from it, leading to a
result close to the curves in the Figure.

\begin{figure}

\hskip 1.1 in  \epsfysize= 2.5 in   \epsfbox{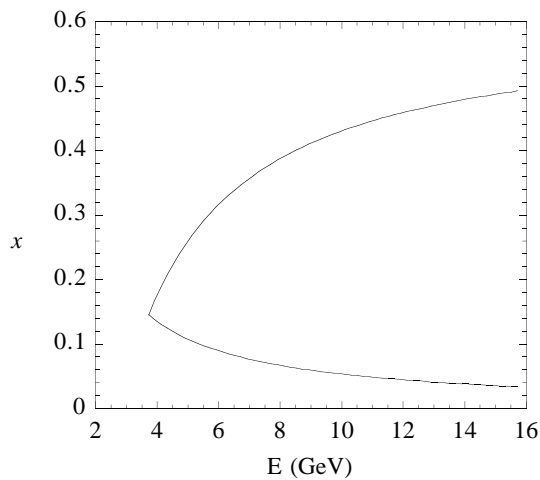}

\caption{Scaling window for $\alpha = 1.4$.}
\label{window}

\end{figure}

How does energy help?  Our analysis has used scaling for the
neutron structure function.  This requires that, at least, 
$Q^2 > 1$ GeV$^2$ and $W$ (the total hadronic mass of the
material coming from the stuck neutron) $> 2$ GeV.  One of
these requirements sets a lower limit on $x$, the other sets an
upper limit.  The allowed values of $x$ are between the two
curves in Fig~\ref{window}~\cite{lassila}.  For incoming electron
energy 4 GeV, the allowed range of $x$ is tiny,  but for 8 or 10 GeV
the range of $x$ is quite enough to see a distinction between
the curves and the flat line in the previous Figure.  


\subsection{An exotic use of TJNAF}


A difficult to observe but interesting CP violating decay is 
\begin{equation}
K_L \rightarrow \pi^0 \nu \bar \nu .
\end{equation}

\noindent The final state is all neutral and two of the
particles are essentially impossible to see, but the motivation
of testing extensions of the standard model and of finding
direct CP violation is strong.

By way of reminder, the long lived neutral kaon is
\begin{equation}
K_L = K_2 + \varepsilon K_1
\end{equation}

\noindent where $K_2$ is the CP odd combination of $K^0$ and 
$\bar K^0$---which cannot decay into two pions by a CP
conserving force---and $K_1$ is CP even.  CP violation has been
seen only in the two pion and semileptonic decays of the $K_L$
(i.e., the $\pi^0\pi^0$, $\pi^+\pi^-$, and $\pi^\pm e^\mp \nu$
modes) and what is seen is compatible with indirect CP
violation, meaning that the CP violation occurs only through
the $K_1$ admixture, follow by a normal CP conserving decay.

\begin{figure}

\hskip 1.3 in  \epsfysize= 1 in   \epsfbox{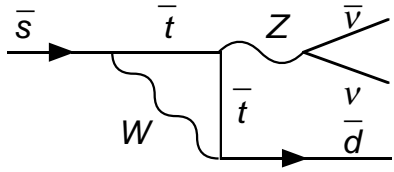}

\caption{Possibility for decay 
$K_L \rightarrow \pi^0 \nu \bar \nu$ in the standard model.}
\label{weak}

\end{figure}

The decay $K_L \rightarrow \pi^0 \nu \bar \nu$ is 100\% CP
violating, and is rare~\cite{buras}.  The decay requires flavor
changing neutral currents, and it can occur in the standard model only
by second order weak interactions such as illustrated in
Fig.~\ref{weak}.  The calculated branching ratio in the
standard model is a few times $10^{-11}$~\cite{buchalla}.


\begin{figure}

\hskip .7 in  \epsfysize= 2.5 in   \epsfbox{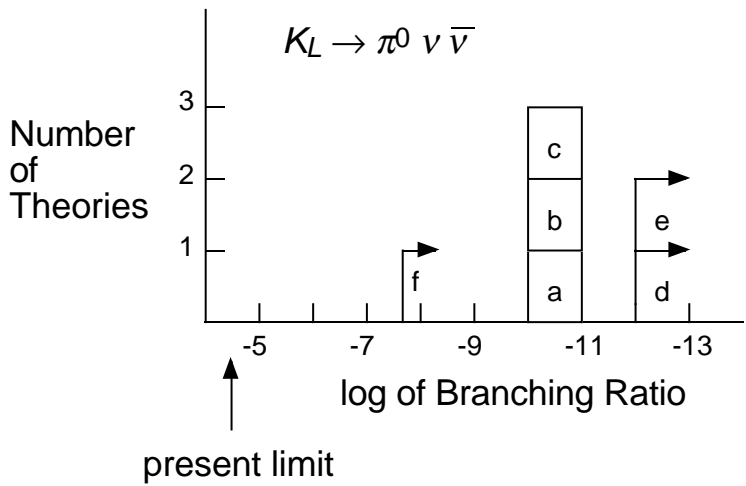}

\caption{Various calculated results for branching ratios of 
$K_L \rightarrow \pi^0 \nu \bar \nu$; (a) standard
model, (b) supersymmetry with R-parity conserved, (c) extra
Higgs doublets with CP violation still in the CKM matrix, (d)
extra Higgs doublets with CP violation only spontaneous, (e)
Weinberg model, and (f) supersymmetry with R-parity
violation.}
\label{results}

\end{figure}

Different extensions of the standard model give different
results for the $K_L \rightarrow \pi^0 \nu \bar \nu$, and the
calculated results known to me are summarized in
Fig.~\ref{results}~\cite{dorata}.  All are below the present
experimental limit.  One, which involves supersymmetry with R-parity
violation~\cite{roy}, may be reachable under the design goals of
presently running experiments.


Could Jefferson Lab actually measure the decay at the level
predicted by the standard model?  The question is under
discussion~\cite{ito}. (It is also under discussion at
Brookhaven National Lab.)  The signal that will be seen is just two
photons, from the decay of the $\pi^0$ and which together have the
mass of the
$\pi^0$, with the neutrinos unseen. The dineutrino mass, if it
can be inferred, will sometimes also have the mass of a
$\pi^0$, but generally not.  High efficiency detectors are
clearly needed, so that things are not unseen merely because of
falling through cracks in the detector.  Dangerous backgrounds
come from the decays 
$K_L \rightarrow \pi^0 \gamma \gamma$ (which occurs at a part
in $10^6$) and $K_L \rightarrow \pi^0 \pi^0$ (one in $10^3$
decays), with the detectors only picking up two of the four
possible photons.  

The time structure of the CEBAF beam may help.  The first word
of the lab's old name is not correct if one can examine the beam
with sufficient time resolution.  There are sharp pulses every
2/3 nanosecond, and there is good control over the pulses, so
that it is possible to leave spaces empty and deliver a sharp
pulse once every (say) 20 ns.  Then we know when the $K_L$ was
formed, and since it has roughly 50 ns lifetime, one can
measure its time of flight and hence its velocity and momentum,
allowing the missing mass to be determined.  If the missing
mass is the mass of a $\pi^0$, cast the event aside.  What
remains will be (one may hope) mainly $\pi^0 \nu \bar \nu$.

This is not an easy experiment, but it interesting just to
think that it may be possible.

\section{The End}

A $9 \pm 1$ GeV electron beam allows some exciting results to
be obtained, and may happen at the Jefferson Laboratory.  One
category of experiments is of the ``more of the same'' variety,
but the extra reach in, for example, $Q^2$ makes them greatly
more interesting.  In addition a number of new initiatives
become possible.

\section{Acknowledgments}

I thank the organizers of this workshop for their work that made this a
pleasant and productive meeting, and to all those who supplied me with
information for this talk.  I also thank the NSF (USA) for support under
grant PHY-9600415.

%
%
\end{document}